\documentclass{nature}
\usepackage{amsmath}
\usepackage{amsfonts}
\usepackage{float}
\usepackage{amssymb}
\usepackage{color,soul}
\usepackage[left]{lineno}
\usepackage{soul,color}
\usepackage{graphicx}
\usepackage{caption}
\usepackage{subcaption}
\usepackage{multirow}
\usepackage{booktabs}
\usepackage[symbol]{footmisc}
\usepackage{indentfirst}

\usepackage[hyphens]{url}
\usepackage[hidelinks]{hyperref}
\hypersetup{breaklinks=true}

\title{High Accuracy Tumor Diagnoses and Benchmarking of Hematoxylin and Eosin Stained Prostate Core Biopsy Images Generated by Explainable Deep Neural Networks}

\author{Aman Rana$^{1\P}$, Alarice Lowe$^{2\P}$, Marie Lithgow$^3$, Katharine Horback$^2$, Tyler Janovitz$^2$,\\ Annacarolina Da Silva$^2$, Harrison Tsai$^2$, Vignesh Shanmugam$^2$, Hyung-Jin Yoon$^{1}$ \& Pratik Shah$^{1}$\thanks{
Correspondence to Dr. Pratik Shah (pratiks@mit.edu) \\ \indent
$^{\P}$ Equal contributions}}

\begin{document}
\maketitle

\begin{affiliations}
 \item Program in Media Arts and Sciences, The MIT Media Lab, Massachusetts Institute \\ of Technology, Cambridge, MA 02139, USA Email: \\
 \{arana, hyungjin, pratiks\}@mit.edu \vspace{0.3cm}
 
 \item Brigham and Women's Hospital, Harvard Medical School, Boston, MA, 02115, USA \\
 \{alowe, adasilva, hktsai, vshanmugam\}@bwh.harvard.edu \\
 \{khorback, tjanovitz\}@partners.org
 \vspace{0.3cm}
 
 \item Boston University School of Medicine, VA Boston Healthcare, West Roxbury MA 02132, USA \\ marie.lithgow@va.gov
\end{affiliations}

\begin{abstract}
ABSTRACT: Histopathological diagnoses of tumors in tissue biopsy after Hematoxylin and Eosin (H\&E) staining is the gold standard for oncology care. H\&E staining is slow and uses dyes, reagents and precious tissue samples that cannot be reused.
Thousands of native nonstained RGB Whole Slide Image (RWSI) patches of prostate core tissue biopsies were registered with their H\&E stained versions.
Conditional Generative Adversarial Neural Networks (cGANs) that automate conversion of native nonstained RWSI to computational H\&E stained images were then trained.
High similarities between computational and H\&E dye stained images with Structural Similarity Index (SSIM) 0.902, Pearsons Correlation Coefficient (CC) 0.962 and Peak Signal to Noise Ratio (PSNR) 22.821 dB were calculated.
A second cGAN performed accurate computational destaining of H\&E dye stained images back to their native nonstained form with SSIM 0.9, CC 0.963 and PSNR 25.646 dB.
A single-blind study computed more than 95\% pixel-by-pixel overlap between prostate tumor annotations on computationally stained images, provided by five-board certified MD pathologists, with those on H\&E dye stained counterparts. We report the first visualization and explanation of neural network kernel activation maps during H\&E staining and destaining of RGB images by cGANs. High similarities between kernel activation maps of computational and H\&E stained images (Mean-Squared Errors $<$0.0005) provide additional mathematical and mechanistic validation of the staining system. Our neural network framework thus is automated, explainable and performs high precision H\&E staining and destaining of low cost native RGB images, and is computer vision and physician authenticated for rapid and accurate tumor diagnoses.
\end{abstract}

\section{INTRODUCTION}
Cancer is the second leading cause of death in the United States~\cite{n1}. An estimated 164,690 American men were diagnosed with prostate cancer and 29,430 succumbed to the disease in 2018~\cite{n1}. Survival rate for people with localized prostate cancer is above 98\%, which drops to 30\% when cancer spreads to other parts of the body such as distant lymph nodes, bones or other organs~\cite{n1}. This drop in survival rate can be prevented with early diagnosis. The current gold standard for prostate cancer diagnosis uses dye staining of core biopsy tissue and subsequent microscopic histopathologic examination by trained pathologists. Hematoxylin and Eosin (H\&E) is the most widely used dye staining method that leverages interactions of hematoxylin and eosin dyes with tissues for visualization~\cite{n2}. Everyday up to three million slides are stained with this technique. Microscopic diagnosis of tumors using H\&E stained biopsy slides present challenges such as inconsistencies introduced during tissue preparation and staining, human errors and also requires significant processing time, imaging systems and procedural costs~\cite{n3}. Other key challenges include sampling time, limited tissue that can be stained due to time and cost involved consequently resulting in evaluation of only three 4 $\mu m$ sections of tissue to represent a 1 $mm$ diameter core. Irreversible dye staining of tissues leads to loss of precious biopsy samples that are no longer available for biomarker testing. Automated, low-cost and rapid generative algorithms and methods that can convert native nonstained whole slide images to computationally H\&E stained versions with high precision can be transformative by benefitting patients, physicians and to reduce errors and costs.

Whole-slide pathology images are United States Food and Drug Administration (FDA) approved~\cite{n4} for cancer diagnosis, and can rapidly be integrated into machine learning and AI algorithms for automatic detection of cellular and morphological structures to tumors and virtual staining~\cite{n5}.  
Studies testing operational feasibility and validation of results obtained by generative models and machine learning algorithms in controlled clinical trials or hospital studies with whole-slide pathology images do not exist, consequently precluding clinical deployment of these systems. As examples, previously reported approaches for automated staining of tissue biopsy using partial images or patches have constraints such as, a) requirement for prestaining tissues prior to excitation with specific wavelengths of UV radiation; b) acquisition of specialized hyperspectral, fluorophore tagged, multispectral images using costly systems; c) staining only few cellular components with low accuracy and limited color spectrum; d) significant loss of information in the stained images; e) limited clinical validation using coarse diagnosis from synthetic images; f) lack of computer vision and image processing methods for benchmarking quality of generated images; and g) no explanation of mechanisms, specifically when neural networks are used, for virtual staining.

We previously communicated convolutional neural networks for learning associations between expert annotations of disease and fluorescent biomarkers manifested on RGB images and their complementary non-fluorescent pixels found on standard white light images~\cite{yauney2017convolutional}.
Subsequently, we communicated Conditional Generative Adversarial Neural Networks (cGANs) that accept native nonstained prostate core biopsy RGB Whole Slide Autofluoroscence Images (RWSI) and computationally H\&E stain them by learning hierarchical non-linear mappings between image pairs before and after H\&E dye staining~\cite{n6}.
Another destaining model converted RWSI of H\&E dye stained prostate core biopsies into their native nonstained form~\cite{n6}.
In this work, we report several novel mechanistic insights and methods to facilitate clinical deployment and regulatory evaluations of these systems.
Specifically, large and diverse training datasets of images of deparaffinized prostate core biopsy RWSI from patients with different grades of tumor were used to, a) train high fidelity, explainable and automated computational staining and destaining algorithms that learn mappings between naturally autofluorescent pixels~\cite{n7} of nonstained cellular organelles and their stained counterparts; b) devise robust loss function for our machine learning algorithms to preserve tissue structure; c) establish that our H\&E staining neural network models generalize to  accurately stain previously unseen images acquired from patients and tumor grades not part of training data; d) generate neural activation maps to provide first instance of  explainability and mechanisms used by cGANs models for H\&E staining and destaining; e) establish computer vision analytics to benchmark the quality of generated images; and f) validate computationally stained images for prostate tumor diagnoses with multiple MD pathologists for clinical deployment (\textbf{Figure 1}).
By describing explainable algorithms that can consistently, rapidly and accurately computationally stain and destain tissue biopsy RWSI, this study generates actionable evidence for clinical deployment, real-world and regulatory evaluations for virtually stained H\&E images and contributes to the fields of clinical development, computer science, oncology and digital histopathology.

\section{RESULTS}
\subsection{Patient metadata:}
Thirty-eight patients (mean age 66.2 years) consisting of White, African American, Hispanic/Latino, and Asian men provided forty-six core biopsy samples. Of these, nine patients had known prostate cancer diagnosis and were undergoing active surveillance. Eighteen patients underwent subsequent prostatectomy and the remaining were either healthy or undergoing prostate cancer treatment at Brigham and Women’s Hospital. Each biopsy sample contained one to six cores of tissue. Zero to 100\% of each tissue core contained prostatic adenocarcinoma of various Gleason grades. Samples were enriched for higher-grade tumors (Gleason grade 4 and 5). 

\subsection{Quantitative evaluation of computationally stained and destained images:}
Computationally H\&E stained whole slide images were compared pixel-by-pixel to corresponding H\&E dye stained images (\textbf{Table 1}). Structural Similarity Index (SSIM), Peak Signal to Noise Ratio (PSNR)-calculated usually in logarithmic (dB) and Pearsons Correlation Coefficient (PCC) were used as quality measures of computationally stained images with H\&E dye stained images regarded as ground truth~\cite{n8,n9}. Average SSIM of 0.902 (max=1), and PSNR of 22.821 dB were calculated, indicating high accuracy of computational H\&E staining of test images (\textbf{Table 1}). High PCC accuracy scores (81.8\% of patches with PCC $\geq$0.7 and 39.4\% patches with PCC $\geq$0.8) indicate that computationally stained patches matched H\&E dye stained patches at a pixel level (\textbf{Table 1}).

Comparison of Red (R), Green (G) and Blue (B) color channels pixel intensities between native nonstained and computationally stained images (-42px \textbf{Table 2}: U$\_$C), and those between native nonstained and H\&E dye stained images (-44px \textbf{Table 2}: U$\_$H) show that computationally stained images had mean intensity difference of only 2px (\textbf{Table 2}: H$\_$C). Similar low differences (underlined) were observed after comparing individual color channels: R (\textbf{Supplementary Table\footnote[1]{Supplementary tables are in Supplementary Material.} (ST) 1}:U$\_$C=-58px, U$\_$H= -58px, \underline{H$\_$C=0px}), G (\textbf{ST 2}: U$\_$C=-6px, U$\_$H= -8px, \underline{H$\_$C=2px}) and B (\textbf{ST 3}: U$\_$C=-62px, U$\_$H= -65px, \underline{H$\_$C=3px}) between ground truth H\&E dye and computationally stained images. 

Prostate core biopsy H\&E dye stained images were computationally destained and compared to native nonstained images as described above. Average PCC, SSIM and PSNR across our test images after destaining were 0.9, 0.963 and 25.646 dB respectively (\textbf{Table 1}), thus showing high similarities with native ground truth nonstained images. RGB pixel intensities between computationally destained and H\&E dye stained images (47px \textbf{Table 2}: H$\_$D), and native nonstained and H\&E dye stained images  (44px, \textbf{Table 2}: H$\_$U) also indicated that computationally destained and ground truth nonstained images only had 3px difference in their overall intensities (\textbf{Table 2}: D$\_$U).  These results indicate high fidelity of learning, reproducing and erasing of multi-chromatic information by computational H\&E staining and destaining algorithms. Average change in pixel intensities in the R and B channel was higher compared to the G channel because H\&E dye predominantly consists of blue and red/pink colors.

\subsection{Analyses of physician annotations:}
United States medical board certified/trained pathologists examined dye stained or computationally H\&E stained images generated by neural networks (additional details in methods section) for prostate tumor diagnoses. Both sets of physicians were not told the source of images provided to them or the details of the study. Another independent pathologist ratified diagnoses and tumor labels provided by both sets of reviewers and compared results to patient records (\textbf{Figure 1}). Intersection over union indicating agreements or disagreements between pathologists examining the same set of images (intra-IOU) was calculated by pixel-by-pixel comparisons of their tumor and non-tumor annotations (\textbf{ST 4}). Pathologists examining H\&E dye stained images had high average intra-IoU agreement scores (IoU\textsubscript{H\&E\_dye\_stained}=0.81) for diagnosing any tumors (\textbf{ST 4}). Pathologists examining computationally H\&E stained images also had high and comparable average intra-IoU agreement scores (IoU\textsubscript{computationally\_stained}=0.77) for diagnosing any tumor (\textbf{ST 4}). These results indicated high internal consistency in clinical diagnosis provided by each set of pathologists on their respective images. Furthermore, tumor diagnoses using computationally stained images did not have an impact on rater sensitivity or specificity while detecting tumors.

Tumor labels provided by two sets of physicians in our single-blind study on ground truth H\&E dye stained images or computationally stained images were then compared using inter-IoU agreement score metric~\cite{n10} (\textbf{Table 3}). An overall inter-IoU score of 0.79 was calculated for any tumor diagnoses. An average inter-IoU agreement score of 0.70 was calculated for Gleason grade 3 labels, while scores of 0.73 and 0.64 were calculated for Gleason grade 4 and 5 labels (\textbf{Table 3}). Average inter-IoU agreement score of 0.90 was calculated for annotations of healthy areas in the tissue where no tumors were found on images (\textbf{Table 3}). For example images in \textbf{Supplementary Figure\footnote{Supplementary figures are in Supplementary Material.} (SF) 3 and 12} were entirely benign and both sets of reviewers did not annotate tumor labels on these images. These results indicate that our trained machine learning models can accurately generate both tumor and non-tumor signatures via H\&E staining. And physician raters showed concordance and comparable sensitivity and specificity in diagnoses made using H\&E dye stained with those made by using computationally stained images.

\subsection{Clinical evaluations of computationally stained images:}
\textbf{Figure 2} shows representative input nonstained image patches in row (a) that had Gleason grade 3 (columns I, II) or 4 (columns III, IV) tumors or were benign (column V), and their computational H\&E staining (row c) and accuracy calculated using annotations by multiple physicians (row d). Tissue morphology in computationally stained patches (\textbf{Figure 2: row c}) matches closely with H\&E dye stained patches (\textbf{Figure 2: row b}). Patch c-I successfully generated a benign area along with tumor signature (as indicated by arrows) and confirmed in \textbf{Figure 2: row d-I}. Computationally stained patches (\textbf{Figure 2: row c}) retain appearance of benign and malignant glands and stroma seen in H\&E dye stained patches (\textbf{Figure 2: row b}). Patch b-III also contains edge/crush artifact (arrowheads) that is preserved in computationally stained image (\textbf{Figure 2: row c-III}). \textbf{Figure 2: row d} shows same patches with color-coded areas of agreement and disagreement between the labels provided on H\&E dye stained images and computationally stained RWSI. It is evident that the computationally H\&E stained patches represent tumor signatures with high accuracy and pathologists are able to correctly identify tumor. Majority of observed disagreements between raters did not represent misidentification of glands as benign or malignant. Instead, they show differences in rater annotation at borders of tumor labels, mainly due to differences in labeling style with some raters providing course labels and others annotating detailed labels (\textbf{Figure 2: row d-III, arrows}), or biopsy edges, as some raters chose to score partial/crushed glands at the periphery of samples and others did not (\textbf{Figure 2: row d-III, arrowheads}).

Reconstructed computationally stained images shown in \textbf{SF 1b and 2b} (used for validation of the trained neural network) morphologically represented benign and malignant glands and stroma well enough to be consistently identified by pathologists (\textbf{SF 1c and 2c}) when compared with corresponding H\&E dye stained images (\textbf{SF 1a and 2a}). A vast majority of tumor also showed annotator agreement. In some instances, ``atypical'' glands that were morphologically indeterminate for malignancy led to interpretative discrepancies however showed preserved morphology in the computationally stained images (e.g. \textbf{arrows in SF 1a and 1b}). Ground truth nonstained (\textbf{SF 1d and 2d}) and corresponding computationally destained images (\textbf{SF 1e and 2e}) are also shown for comparison.

\textbf{SF 4b, 6b and 7b}, show the most reported areas of disagreement many of which are attributed to atypical glands that were hard to categorize on both images but were well represented on the computer-generated images (\textbf{SF 4c, 6c and 7c}). \textbf{SF 5} shows the uncommon Gleason pattern 5 tumors with comedo necrosis (\textbf{SF 5a, arrow}). The morphology of the tumor glands is well maintained (\textbf{SF 5b, arrowheads}), but the comedo necrosis is not visualized (\textbf{SF 5b, arrow}). The dye-stained image in \textbf{SF 7a} contains an infrequently encountered scenario (indicated by an arrow), the presence of rare malignant glands that are not well visualized on the computationally stained image (\textbf{SF 7b, arrow}). Despite this altered appearance, there was no impact on clinical diagnosis as the blinded reviewers scored these areas as tumor. Some glands are poorly formed on both the dye stained and the computationally stained image (\textbf{SF 8a, 8b, arrows}), leading to disagreement between raters, even though the computationally stained image were identical to dye stained image.

Images shown in \textbf{SF 9} presented a challenging labeling exercise where tumor cell cytoplasm was very pale and did not show significant contrast to the background stroma in the dye-stained image (\textbf{SF 9a}). This cytoplasmic pallor was also well preserved in the computationally stained image (\textbf{SF 9b}). Despite this, appearance of the nuclei and the slight difference in cytoplasmic texture made the tumor identifiable in both images (\textbf{SF 9c and 9d}). The computationally stained images shown in \textbf{SF 10b, 11b and 13b} were well represented. Majority of the disagreement in these images arose due to tumor/non-tumor boundary and biopsy edge issues. Validation images in \textbf{SF 11b and 13b} illustrated additional high-quality examples of preserved morphology generated by the computationally staining algorithm, which confirmed accurate matching with dye stained images in benign conditions. Non-necrotizing granulomas, marked chronic inflammation, reactive stromal changes and proteinaceous debris were all morphologically identifiable in the computational stained images (\textbf{SF 11c}). Pathologists unanimously scored the matched H\&E dye stained and computationally stained images shown in \textbf{SF 3 and 12} as benign.

\textbf{SF 14} shows additional examples of indeterminate atypical glands (columns I and II) and tumor with edge/crush artifact (columns III and IV) that are well preserved on the computationally generated images but differentially designated as tumor or non-tumor by raters. \textbf{SF 14} column V shows non-necrotizing granulomas (arrows in c-V), which represent the only example of this feature in our training and validation sets. Despite not being encountered before, the morphology of the granulomas was relatively well maintained, resulting in the correct identification by raters and categorization as non-tumor.

\subsection{Comparison with patient records:}
A vast majority of the diagnoses rendered using computationally stained images agreed with the corresponding initial clinical diagnosis reported in Electronic Health Records (EHR) (\textbf{Table 4}), supporting the validity of the generated images for tumor detection and diagnoses. Majority study cases showed identical tumor fractions and Gleason grading as previously reported. After expert re-review of the original slides and additional evaluation by immunohistochemistry, the original EHR diagnosis was overturned in two cases, resulting in two additional cases of agreement. Pathologists reviewing computer-generated core 11 were able to better identify the presence of rare glands of Gleason grade 3 tumors than those who had rendered the original EHR diagnosis of benign (\textbf{SF 11 marked blue/green}). Microscopic re-review of the original glass slide confirmed that it indeed had a tiny focus of grade 3 tumor that was overlooked at the time of the original diagnosis. Subsequent immunohistochemical analysis revealed the absence of basal cells around the glands in question, confirming the diagnosis of carcinoma made during this study and revealing the diagnosis conferred on the computationally generated images to be correct. \textbf{SF 2} was the only study biopsy that showed a significant difference in tumor fraction, as this study reported 50\% tumor fraction and the original EHR report was 90\%. Re-review of the original glass slide again showed this study fraction to be more accurate than the original diagnosis (\textbf{SF 2}). Otherwise, the tumor fraction identified in all the computationally generated images approximated the fraction reported in the EHR for all images as evident from \textbf{Table 4}.

None of the differences between EHR and computationally generated H\&E diagnosis were clinically significant with regard to treatment decisions. A difference in grade of tumor was identified in a minor component of computationally stained images (\textbf{SF 4, 7 and 13}). The small foci of higher or lower grade tumor identified in computationally stained images (\textbf{SF 4c, 7c, and 13c}), which were not reported at the time of original diagnosis, comprised a very small fraction of tumor volume. These were often associated with diagnostically indeterminate questions (e.g. whether a gland represented a rare focus of grade 4 tumor or if it was tangential sectioning of grade 3 tumor), and were not clinically significant in the context of the patient's known tumor at the time of original EHR reported diagnosis.

\subsection{Analysis and explanation of neural network activation maps:}
Neural activation maps of trained staining and destaining cGAN models were analyzed after feeding healthy (shown in \textbf{Figure 2 row a-V}), Gleason grade 3 (\textbf{Figure 2 row a-I}), 4 (\textbf{Figure 2 row a-IV}) or 5 images patches. Majority of previous research in explaining neural networks trained for image classification correlate and classify convolutional neural network activation maps kernels back to input images being analyzed~\cite{yosinski2015understanding}. In this work we do not use a classification approach to identify image features, but rather perform pixel-by-pixel visualization, explanation and intensity ranking ($>$200 value) of various cGAN kernels to create an activation map of a particular nonstained image patch (healthy or with a particular Gleason tumor grade) as it passes through each network layer while getting stained (\textbf{Figure 3}). To our knowledge, no prior study (other than the research in our laboratory~\cite{n6}) has described computational destaining of H\&E dye stained images to revert them to their native nonstained form for virtual staining with other dyes or for conversion to immunohistochemistry feature space. Computationally destained images in this study were also used to evaluate the activation profiles of trained neural network H\&E staining models (\textbf{Figure 3}).

We demonstrate and compare presence of unique low and high-level features in input images (computational vs. dye stained for e.g.) that activate neurons and feature maps in the cGAN generator network (\textbf{Figures 4, 5 and 6}). For example, initial layers of the convolutional layers in the generator detect low-level features such tissue geometry, edges, corners, shapes and a few changes in color (\textbf{Figures 4 and 5, Layers L1, L2 \& L3, panels I, III \& V}). We see well-demarcated boundaries between tissues and background (\textbf{Figure 4 and 5, Layers L1, L2 and L3, panel I}) and gross distinctions between glands and stroma are suggested (\textbf{Figure 4, Layer L2, panel I}) or are well defined (\textbf{Figure 5, Layer L1, panel I}). A high amount of activation in background pixels of the image (not containing tissue) was observed in majority of panels in layers L1 to L5 (\textbf{Figures 4 and 5}). Panel I in layers L1 and L3 in \textbf{Figures 4 and 5} show an excellent example of activation of neural network kernel switching between focusing most significantly on the background (Panel I in layer L1) vs. another kernel almost exclusively being activated in response to tissue (Panel I in layer L3).  Kernels of initial layers of trained models thus help with differentiating tissue from background and morphological tasks to define higher order anatomical structures (\textbf{Figures 4 and 5}). 

The later convolutional layers leverage previously learned low level features and ability to differentiate tissue from background with fine-grained structures such as anatomical arrangement of nuclei and tumor signatures (\textbf{Figures 4 and 5, Layers L17, L18 and L19}).  For example, maps in layers L17 in \textbf{Figures 4 and 5} illustrate that the neural network activation profile is significantly minimal for background, morphology and shapes but rather responsive to fine grained microscopic features contained in the tissues. Significant activation is also seen in boundaries separating stroma from glands, whether benign or malignant (\textbf{Figure 5, Layer L17, L18 and L19}). \textbf{SF 16 and 17} show additional examples activation maps of images of prostate core biopsy with Gleason grade 4 and 5 tumors which demonstrate similar activation patterns, suggesting a conserved mechanism is used by trained generator to perform computational staining across tumor grades.

We compared kernel activation maps of all 448-validation image patches used to test our trained staining and destaining machine learning models with corresponding ground truth dye stained and native nonstained images (\textbf{Figure 6}). Mean-Squared Error (MSE) was calculated by comparing activation maps generated by each of the 19 neural network layers in response to pairs of images being evaluated. Translucent blue lines show the MSE for all the 448 input patches. The red line shows that the average MSE was very low and max average NMSE is less than 0.0005. \textbf{Figure 6a} shows that native nonstained images and corresponding computationally destained input images activated our trained computational staining neural network layers with high similarities while being stained. We also calculated similar low MSE (\textbf{Figure 6b}) in activation maps when computationally stained and corresponding H\&E dye stained images were fed to our trained computational destaining model. MSE was low for 1\textsuperscript{st} layer, increases for 2\textsuperscript{nd} layer and then decreases for the remaining layers. The MSE peaks for layers 3, 10 and 17 were slightly higher suggesting unique activation patterns were more prevalent in kernels residing in these layers. These results, in unification with our detailed SSIM, PSNR, CC and physician validation, provide significant evidence of high quality of computationally stained and destained images, with consequent high sensitivity and specificity in diagnosing tumors using them.

\section{DISCUSSION}
A vast majority of surgical and medical treatments for cancer, including chemotherapy, endocrine therapy, and immunotherapy are dictated by histopathologic examination and diagnosis. Increase in use of core biopsies for diagnosis, in place of larger surgical biopsies, has resulted in significant decrease in the volume of tumor available for performing an ever-increasing battery of biomarker testing for diagnostic, prognostic, and predictive information. If a new process to obtain an instant and accurate computational H\&E staining of native nonstained WSI of prostate tissue is developed, it will accelerate process of conventional histopathology and save precious tissue samples.

Computationally stained and destained images reported in this study were evaluated by multiple image analytics and matched ground truth images with high similarity (\textbf{Table 1}). MSE compares the true pixel values of H\&E dye stained images to computationally stained images and is inversely correlated to PSNR~\cite{n8}. Thus, higher the PSNR, the better computationally stained image has been reconstructed to match the original image and the superior the H\&E staining or destaining algorithm. The main limitation of PSNR is that it relies strictly on numeric pixel comparison and does not account for biological factors of the human vision system that detect macro structures~\cite{n8}. Unlike PSNR, SSIM is based on visible structures in the image with the notion that pixels have strong inter-dependencies especially when they are spatially close~\cite{n9}. These dependencies carry important information about the structure of the objects such as tumors, stromas, and glands and other morphological tissue feature in the H\&E images. Whereas PCC test is performed by randomly scrambling the blocks of pixels (instead of individual pixels, because each pixel's intensity is correlated with its neighboring pixels) in H\&E dye stained image, and then measuring the correlation of this image with the computationally stained image~\cite{n12}. Taken together, high quality of the computationally stained and destained images calculated using MSE, PSNR, SSIM and CC cumulatively provided comprehensive and stringent evaluation of their macroscopic and microscopic suitability for clinical deployment.

Evaluation by trained pathologists showed tumorous and healthy tissues were morphologically well represented in majority of the computationally stained images with high accuracy (\textbf{Figure 2} and \textbf{SF 1-13}). The glands and stroma of benign prostatic tissue and carcinoma were identifiable, showing preserved architectural features (location and shape of the glands), defined gland/stromal interface, and cytology (including location and appearance of the nuclei and nucleoli, if present). A majority of the differences in annotations (such as those seem in \textbf{SF 4 and 13}) were observed either on the tumor/non-tumor interface/boundary or the biopsy boundary. This can be attributed to labeling style of individual raters, where some raters gave detailed labels while the others gave course labels and some chose to label crushed glands at the periphery of the tissue. Previous studies report that human readers show substantial variability and lower average performance than computer algorithm in terms of tumor segmentations~\cite{n13}.  Similar limitation of using a human reader panel to establish a reference standard for evaluation of computer algorithms may have impacted this study. In validation images, presence of morphologically ambiguous glands, a known histopathologic dilemma that clinically requires additional work up for confident diagnosis, also led to differing labels between raters as they were asked to categorize each gland as benign or malignant without assistance from supplemental studies. In most cases (\textbf{SF 8}) these ambiguous cases were well represented in the computationally stained images (arrows in \textbf{SF 8b}), but led to labeling differences due to the ambiguity of these said regions of interest (ROI).
 
Small difference calculated by PSNR, SSIM, CC (\textbf{Table 1}), independent of the human raters, may also be in-part due to registration differences in small out-of-focus areas during WSI~\cite{n14}. Input image pairs (nonstained and H\&E stained) used for training in our work were corrected for differences in field of view, illumination and focal planes but may still have minor variances. These small variances in computationally stained images though had no impact on overall clinical assessments. Color variations in digital slides may arise due to differences in staining reagents, thickness of tissue sections and staining protocols and can negatively impact clinical diagnoses~\cite{n5}. We report minimal color variation across our 13 computationally stained H\&E images as seen by their uniform overall RGB and individual R, G and B channel intensity values which often match training images (\textbf{Table 2, ST 1, 2 and 3}). Physician raters in the study did not report difficulty in reading colors of nuclei, glands, cells and tumors in computationally stained images, which was ratified by an additional independent pathologist. Thus, the trained neural network model reproduces a consistent and normalized color hue from the vast training dataset that does not impact clinical decision making from computational images. And the subsequent absence of false-positive errors in healthy tissue cores of patients illustrates the fine grain reproduction of our computationally stained and destained images.

We were also pleased to find high concordances between diagnosis made using the computationally stained images in this study and patient EHR (\textbf{Table 4}). We in fact found two instances where the diagnoses made using computationally stained images overturned the initial EHR findings. In both cases, additional laboratory tests and clinical workups were performed to confirm our findings. These results demonstrated that raters and the tumor diagnoses performed using computationally stained WSI used in our study matched or exceeded the initial microscopic diagnosis performed using H\&E stained tissue slides after prostate biopsy extraction.
 
Virtual staining of histopathology slide images has been reported using approaches with signals that require long detection times~\cite{n20}, dye staining of nonstained specimens prior to imaging~\cite{n21}, laser illumination and excitation with specific wavelengths~\cite{n22} and sparse sampling and poor depth resolution~\cite{n21,n23}. Previous virtual staining studies have performed limited analytics~\cite{n24} to benchmark the quality of their virtually stained images. Majority do not perform pixel level comparisons with ground truth images and use small numbers of non-blinded raters who use coarse annotations, without tumor gradations~\cite{n25,n26}. While others report no clinical validation and benchmarking of their results~\cite{n27,n28,n29}. Another recently published study required prestaining tissues with eosin, rhodamine and several other dyes before imaging with multiple UV wavelengths to generate autofluorescence images, which were then evaluated by a single pathologist~\cite{n21}. In summation, these techniques and studies have shown limited spatial resolution to locate small tumors, suffer from auto fluorescence and specular reflections issues, required prestaining or specialized illumination sources, fluorescence scopes and sensors and/or conduct limited image quality or clinical evaluations.

Similarly, previous deep learning research for virtual staining uses specialized illumination sources and does not report robust validation studies on mechanisms to establish computer vision or diagnostic utility of generated images~\cite{n5}. Bayramoglu et. al. virtually stain lung tissue slide multispectral images with a cGAN and achieve a structural similarity (SSIM) of 0.3873 but perform no clinical validation~\cite{n30}. Bulingame et. al. use cGAN to convert H\&E stained pancreas slide RGB images to immunofluorescence images and achieve a SSIM of 0.883, and also do not report clinical validation of generated images~\cite{n31}. Rivenson and coworkers use a fluorescence scope with specialized UV filters to capture various tissue biopsy images and virtually H\&E stain them using a neural network~\cite{n32}.
Results and findings communicated in our study differ from previous deep learning based virtual staining studies in several key aspects. As examples, Rivenson et.al. utilized a wide field fluorescence microscope to image tissue~\cite{n32} vs. the non-fluorescent mode of a FDA cleared and widely available automated slide scanning system to capture images used in our study. A single pathologist compared anatomical features between virtually stained images using coarse labels, and pixel-level comparisons between tumor labels on virtual and ground truth images or concordance with EHR of patients were also not conducted to calculate true and false positive occurrences of tumor diagnoses reported in that study~\cite{n32}. Computational destaining of tissue images and stringent image analytics such as PSNR or CC to benchmark quality of virtually stained images have not been reported in previous deep learning based studies~\cite{n30, n31, n32}. Analysis or visualization of key neural network kernels and image features that get activated during the staining process have not been, and thus precluding mechanistic insights or mathematical validation of previous findings reported in literature~\cite{n30, n31, n32}.

In this study, we communicate trained neural network models that computationally H\&E stain native unlabeled RGB images of prostate core biopsy (acquired without band pass filters or specialized hardware) with anatomical features of prostate and reproduce cancer tumor signatures with high accuracies. Computational pixel-by-pixel analysis and comparisons using PSNR, SSIM and PCC demonstrate high similarities between our computationally stained images and their H\&E dye stained counterparts. Pixel-by-pixel changes in R, G, and B color channels after computational staining and destaining by neural networks match corresponding changes in RGB intensity when native nonstained images are H\&E dye stained in pathology labs vice versa.  Detailed clinical validation in a single blind-study found high inter and intra-rater agreements,  calculated by pixel-by-pixel analyses of tumor labels provided by multiple board certified/trained physicians. Computationally stained images thus accurately represented healthy tissue as well as tumors of different Gleason grades, which were easily detected by human visual perception. Clinical diagnoses made using computationally stained images in our study were consistent with tumor diagnoses reported in EHR. We investigate layers of generator neural networks and calculate activation of kernels during staining of different prostate tumor grades and benign tissue signatures to visualize and explain the process of computational H\&E staining and destaining. Activation maps of our trained neural network models during computational staining or destaining of test images were highly similar to H\&E dye stained or native nonstained images. Thus by visualizing and comparing activation feature maps of kernels of trained models this work also presents the first explainable deep neural network framework for computationally H\&E staining or destaining of native RGB images. 

This study provides framework for generating actionable and explainable evidence for regulatory evaluations prior to conducting controlled clinical trials for establishing efficacy of Artificial Intelligence algorithms as clinical support systems~\cite{n33,n34,shah2019artificial}. We also communicate foundational work for adopting computational H\&E staining methods in clinical environments for enabling saving of time and effort required for manual staining and slide preparation, and more importantly preservation of precious tissue samples which could be used in a targeted fashion for biomarker evaluation.

\section{METHODS}
\subsection{Data collection, transfer and processing of whole slide images:} Partners Human Research Committee (Boston, MA) approved protocol 2014P002435 ***********, after which excess material from prostate core biopsies performed in the course of routine clinical care between 2014 and 2017 at Brigham and Women's Hospital (BWH), Boston, MA, were obtained for this study. Forty-six non-stained and corresponding H\&E dye stained RWSI were collected from 38 patients and imaged at 20x magnification.  Briefly, prostate core biopsy specimens were immediately fixed in 10\% formalin, paraffin embedded, cut into 4-micron thick sections and placed on standard glass slides that were placed in archival storage at room temperature. Deparaffinized nonstained slides were scanned with the Aperio ScanScope XT system (Leica Biosystems, Buffalo Grove, IL) at 20x magnification. Subsequently, slides were stained with H\&E dye on the Agilent Dako Autostainer (Agilent, Santa Clara, CA), and these stained slides were re-scanned on the Aperio ScanScope XT at 20x magnification at Harvard Medical School Tissue Microarray \& Imaging Core. Deidentified data in the form of nonstained and H\&E dye stained images at 20x magnification were analyzed at Massachusetts Institute of Technology. Individual prostate tissue needle core biopsy images from each whole slide image were extracted. Extracted core images were horizontally or vertically rotated to reduce non-tissue pixels. This resulted in 102 high-resolution native nonstained and H\&E dye stained image pairs. 

\subsection{Image registration and processing:} Deparaffinized single core images (henceforth called as nonstained images) and subsequent H\&E dye stained single core images of the same biopsy (henceforth called as H\&E dye stained images) were registered using Photoshop CC software (Adobe Systems, San Jose, CA) and corrected for variances~\cite{n35,n36}. Tissue shearing during the staining procedure resulted in regions that could not be registered that were cropped and discarded.

\subsection{Training and validation datasets:} The registered dataset of images was divided into training (82 image pairs) and validation images (13 image pairs). The final patch based training and validation dataset consisted of 74K (training) and 13.5K (validation) image patches.
Validation and training datasets were balanced to include images from healthy patients as well as patients with different grades of prostate tumors and of each tumor grade. The size of the RWSIs were too large to feed into deep learning networks, therefore each image was cropped into multiple patches of size 1024$\times$1024$\times$3 pixels.

\subsection{Machine learning model architecture:} A cGAN pix2pix based model was trained to learn distribution and mappings between registered images in the training dataset~\cite{n37}. The trained model can accept native nonstained images and generate corresponding computationally H\&E stained images via the learnt feature space. If I$_{u}$ and I$_{s}$ represent the native nonstained and H\&E stained image patches in the training dataset. The generator takes in I$_{u}$ as the input and generates I$_{cs}$, the corresponding computationally stained image patch, as the output. The discriminator analyses the output image I$_{cs}$ and predicts the probability that I$_{cs}$ is real (from the training dataset) or fake (output from generator). While training, the generator learns to create images, which can fool the generator while the discriminator learns to correctly identify the fake images. The generator and the discriminator thus play a {\it min-max} game trying to outlearn each other. A novel PCC term was devised specifically for training our neural network models that was added to the cGAN loss function to improve the quality and enforce tissue structure preservation of computationally stained images. The loss function consisted of the cGAN loss~\cite{n10}, a L1 component and a PCC factor between I$_{s}$ and I$_{cs}$. The PCC term in the loss function help reduce the tiling artifacts in the computationally stained images. The loss equation was :
\begin{align*}
\mathcal{L}_{cGAN}(G, D) &=\mathbb{E}_{x,y}[log D(x, y)] +  \alpha \mathbb{E}_{x,z}[log(1-D(x,\ G(x, z)))]\\
\mathcal{L}_{L1}(G) &= \mathbb{E}_{x,y,z}[\parallel y-G(x,z) \parallel_{1}] \\
\mathcal{L}_{PCC}(G) &= \mathbb{E}_{x,y,z}[PCC(y,\ G(x,z))]
\label{equation_1}
\end{align*}
The final loss function is:
\begin{equation*}
G^{*} =\ arg\ \underset{G}{min}\ \underset{D}{max}\ \mathcal{L}_{cGAN}(G, D)\ +  \lambda \mathcal{L}_{L1}(G) + \gamma \mathcal{L}_{PCC}(G)    
\end{equation*}
where $x$ is the input image, $y$ is the target image and $z$ is the random noise, added as dropout in our work. $\mathcal{L}_{cGAN}(G, D)$ is the cGAN loss function, $\mathcal{L}_{L1}(G)$ is the L1 loss between the output of the generator and the target image, and $\mathcal{L}_{PCC}(G)$ is the proposed term that calculated the Pearson\'s correlation coefficient between the generator output and target image. $\alpha=1$, $\lambda=100$ and $\gamma=10$ gave best results. After training, the model accepted unseen native nonstained image patches and generated computationally H\&E stained images patches.

\subsection{Training and validation of machine learning models:} Two machine learning models were trained a staining model that generates computationally H\&E-stained RWSI patches using previously unseen non-stained and native RWSI patches as input, and a destaining model that reverses the process and computationally destains previously unseen H\&E dye-stained RWSI patches. Both models were trained using 74K patches and validated on 13.5K patches.
The discriminator was trained after every single training step for the generator. Both networks were trained for 10 epochs each using Adam optimization~\cite{n31}, and a batch size of one on a NVIDIA GeForce 1080 TI GPU (NVIDIA, Santa Clara, CA) with 12 GB of VRAM and CUDA acceleration to speed up training. One epoch of training (74K training patches) took approximately 16 GPU hours. The patches were randomly flipped and dropout was used to prevent over-fitting and increase generalization capability of the model.

\subsection{Evaluation metrics:} The computationally stained image patches Ics, generated by our model were compared to the H\&E dye patches Is to obtain a quantitative measure of the generated images. PCC, PSNR and SSIM were used to quantify similarities and differences between a given pair of images at a pixel level. The values of PCC and SSIM ranges from 0 to 1, higher values are better. Acceptable values of PSNR for wireless transmission quality loss are considered to be between 20dB to 25dB. Higher PSNR is better. The average and total increase in pixel intensity after computationally staining and destaining was calculated by subtracting the mean pixel intensity of the second image from the first.

\subsection{Clinical validation of computationally H\&E stained RWSI:} Clinical diagnoses by five physicians to investigate, compare and evaluate the efficacy of computationally stained images for tumor diagnoses was conducted. Computationally stained patches from the validation set were used to create reconstructed RWSI images. The corresponding RWSI H\&E dye stained images were used as ground truth examples and also labeled for tumors. A single blind study was conducted for evaluation of generated images in clinical settings for prostate cancer diagnosis. Four board certified/trained expert pathologists provided detailed labels in the forms of free-form outlines encompassing tumors, indicating tumor regions (with grade) and other atypical manifestations on the computationally stained images and H\&E dye stained image. In the first round, two randomly selected pathologists were provided computationally stained images while H\&E dye stained images were given to the other two raters. After a period of four weeks the image sets were swapped between the pathologists, and another round of annotations were conducted. Pathologists annotated images in the form of free hand drawing using the Sedeen Viewer (PathCore Inc., Toronto, Ontario, Canada) on identical notebook computer screens (Dell Computers, Round Rock, TX). By using different colors corresponding to each tumor grade, annotations were classified with tumor grade - Gleason grade 3 (G3), Gleason grade 4 (G4), Gleason grade 5 (G5). A separate comments box was used to note other clinical observations and for anatomical features. The annotations and the associated labels (G3, G4, G5) were extracted from the XML files generated by Sedeen, using the labels and annotations using Python code. ater agreement was calculated in the form of IoU by using the overlap in rater annotations on computationally stained and corresponding H\&E dye stained images~\cite{n10}. A fifth physician was provided computationally stained and H\&E dye stained image pairs leaded in Sedeen to perform qualitative comparisons for histologic structures and features. The final tumor labels on the H\&E dye stained images and corresponding computationally stained images were ratified by an independent clinical pathologist. Accuracies and errors were calculated using pixel-by-pixel overlap in the labels. Color-coded error overlaid validation images were generated visualizing the true positives (green), false positives (red) and false negatives (blue) (\textbf{SF 1-13}).

\subsection{Activation maps:}
Input images containing Gleason grade 3, 4 and 5 signatures were fed into our trained computational staining network to visualize activation maps for each input image. RWSI (full scale RGB images at 20x resolution) were collected/constructed for the following eight image datasets (four pairs) each with 13 images: [Ground] Native nonstained; [Ground] H\&E dye stained; [Predicted reconstructed] Computationally stained; [Predicted reconstructed] Destained (also referred to as predicted destained images). 448 unique patches in each of the eight datasets with no overlap were created for each of dataset and set to size 1024$\times$1024$\times$3 (3 color channels). For each matching patch pair to be fed into computational staining or destaining models, we go linearly over the grid and isolate consolidated activation maps from layer 1 to layer 19 (shown in \textbf{Figure 3}).
A script code ($pix2pix\_activation\_analysis.py$)~\cite{n37} created individual activation maps per layer of the model architecture that are resized to a fixed size (128$\times$128 pixel), and then concatenated together to form a single image per layer of the model architecture. An example concatenated activation map can be seen in \textbf{Figures 4 and 5}. Each image contains 64 activation maps that have been concatenated together. Normalization of the grayscale grid activation maps to 0 to 1 by dividing each pixel by 255 (maximum value of any pixel) was done to dial the MSE value down from 65,025 to 1.0. i.e. the upper bound on the resulting MSE is 1.0.  MSE was calculated using $sklearn.metric’s mean\_squared\_error$ API function. MSE values for all layers (between the matching patch pairs) were collected in a list (array) and plotted on a graph using $matplotlib.pyplot$ library. The y-axis (MSE value) ranges from 0 to 1.0 and the x-axis (layer ID) ranges from 1 to 19 (\textbf{Figure 6}).

\begin{addendum}
 \item[Acknowledgements:] Authors thank Keya Larrel Oliver for technical assistance.
 
 \item[Author contributions:] Alarice Lowe collected data. Aman Rana organized and preprocessed data received from collaborating institution. Aman Rana wrote code, trained, tested and validated machine learning algorithms. Aman Rana, Hyung-Jin Yoon and Pratik Shah analyzed results generated by machine learning algorithms. Marie Lithgow, Katharine Horback, Tyler Janovitz, Annacarolina Da Silva, Harrison Tsai and Vignesh Shanmugam provided tumor annotations on images used in this study and comments on manuscript sections. Pratik Shah, Aman Rana and Alarice Lowe analyzed and interpreted data labeled by expert physicians.  Pratik Shah, Aman Rana, and Hyung-Jin Yoon performed literature search. Aman Rana and Hyung-Jin Yoon made figures and tables. Aman Rana, Alarice Lowe and Pratik Shah wrote  manuscript. Pratik Shah conceived, supervised and directed the research for this study.
 
 \item[Data availability:] Anonymized patient level data and or full dataset will be made available following standard MIT Committee on the Use of Humans as Experimental Subjects and Partners Human Research Committee data sharing protocols.
 
 \item[Competing interests:] The authors declare that they have no competing financial interests.
 
 \item[Correspondence:] Correspondence and requests for materials should be addressed to Dr. Pratik Shah. (email: pratiks@mit.edu).
\end{addendum}
\newpage

\noindent\textbf{Figure 1.} Overview of process of Computational staining and destaining of whole slide prostate core biopsy images with Conditional Generative Adversarial Neural Networks (cGAN) shown on left vs. traditional staining with Hematoxylin and Eosin (H\&E) dyes using physical prostate core tissue biopsy slides (right) and clinical evaluation by multiple physicians of images for tumor diagnosis (left and right) described in this study.

\noindent\textbf{Figure 2.} Representative image patches generated by the computational staining neural network and their comparison with corresponding ground truth Hematoxylin and Eosin (H\&E) dye stained images. Row (a) shows deparafinized native nonstained image patches fed to the neural network. Row (b) shows ground truth H\&E dye stained patches. Focal crush/edge artifact in b-III is indicated by arrowhead. Row (c) shows computationally H\&E stained patches generated by the neural network. Arrows in c-I indicate the two benign glands, all other glands represent tumor. Row (d) shows computationally H\&E stained patches overlaid with colors indicating agreements and disagreements between physician annotations on these images compared to ground truth H\&E dye stained images. Crush/edge effect (arrowheads) and variation in labeling detail by annotators (arrows) are shown in d-III. Green is true positive, blue is false negative and red is false positive.
\newpage

\noindent\textbf{Figure 3.} Visualization and explanation of computational Hematoxylin and Eosin staining process by custom auto-encoder neural network used in this study. Panel (a) Processing of native nonstained prostate core biopsy images as various layers of the encoder and decoder neural networks computationally stain them. The blue boxes represent hidden activation layers of the neural network. Panel (b) A single input native nonstained patch and representative concatenated activation maps (from the corresponding hidden layers in panel (a) of kernels of the decoder neural network as it flows through them, are shown.

\noindent\textbf{Figure 4.} Activation maps of kernels of trained generator neural network model layers after feeding a native nonstained prostate core biopsy image patch without tumor as it gets computationally Hematoxylin and Eosin stained. Rows show top five activation maps from layers L1 - L5 and L16 - L19 arranged in decreasing order of their activations from left to right (columns I-V).

\noindent\textbf{Figure 5.} Activation maps of kernels of trained generator neural network model layers after feeding a native nonstained prostate core biopsy image patch with Gleason grade 3 tumor as it gets computationally Hematoxylin and Eosin stained. Rows show top five activation maps from layers L1 - L5 and L16 - L19 arranged in decreasing order of their activations from left to right (columns I-V).
\newpage

\noindent\textbf{Figure 6.} Comparison of Mean Squared Errors (MSE) between kernel activation maps of pairs of 448 validation image patches generated by the trained neural network models. (a) MSE of ground truth native nonstained and corresponding computationally destained input patch activation maps generated by the trained computational staining model; (b) MSE of computationally Hematoxylin and Eosin (H\&E) stained – and corresponding ground truth H\&E dye stained matching input patch activation maps generated by the trained computational destaining model. Blue lines represent the MSE values for each of the 448 input pairs. Red curve represents  average MSE value at each layer of the generator for all input pairs. The green and orange curves represent the first and third quartile MSE values for all input patch pairs. Lower MSE indicates more accuracy between the activation maps being compared.
\newpage

\begin{table}
\centering
\renewcommand{\arraystretch}{0.6}
\begin{tabular}{ccccccc}
\toprule
\multirow{2}{*}{\textbf{Image}} & \multicolumn{3}{c}{\textbf{Computational staining}} & \multicolumn{3}{c}{\textbf{Computational destaining}} \\
 & \textbf{PCC} & \textbf{SSIM} & \textbf{PSNR} & \textbf{PCC} & \textbf{SSIM} & \textbf{PSNR} \\
\midrule
1 & 0.950 & 0.860 & 20.563 & 0.951 & 0.853 & 23.486 \\
2 & 0.952 & 0.891 & 22.387 & 0.965 & 0.895 & 25.706 \\
3 & 0.949 & 0.860 & 20.683 & 0.964 & 0.866 & 24.871 \\
4 & 0.957 & 0.929 & 22.870 & 0.968 & 0.936 & 27.469 \\
5 & 0.960 & 0.947 & 24.838 & 0.970 & 0.949 & 21.194 \\
6 & 0.955 & 0.914 & 22.903 & 0.957 & 0.914 & 25.285 \\
7 & 0.960 & 0.881 & 22.486 & 0.938 & 0.865 & 21.863 \\
8 & 0.968 & 0.931 & 24.132 & 0.959 & 0.927 & 26.164 \\
9 & 0.978 & 0.890 & 23.411 & 0.968 & 0.874 & 24.165 \\
10 & 0.956 & 0.913 & 23.177 & 0.967 & 0.927 & 27.359 \\
11 & 0.972 & 0.907 & 23.945 & 0.984 & 0.899 & 26.792 \\
12 & 0.975 & 0.902 & 23.200 & 0.963 & 0.899 & 24.957 \\
13 & 0.965 & 0.899 & 22.074 & 0.970 & 0.900 & 26.082 \\
\textbf{MEAN} & \textbf{0.961} & \textbf{0.902} & \textbf{22.821} & \textbf{0.963} & \textbf{0.900} & \textbf{25.646} \\
\bottomrule
\end{tabular}
\caption{Performance analytics of computational staining and destining: Comparison between computationally stained and ground truth Hematoxylin and Eosin (H\&E) dye stained images, and between computationally destained and ground truth native nonstained images. Pearson’s Correlation Coefficient (PCC) of 1.0 means perfect match; Structural Similarity Index (SSIM) of 1.0 means perfect match. Peak Signal to Noise Ratio (PSNR) of 22 dB or more is considered high quality.}

\label{table:pcc_ssim_psnr}
\end{table}


\begin{table}
\centering
\renewcommand{\arraystretch}{0.6}
\begin{tabular}{ccccccc}
\toprule
\multirow{2}{*}{} & \multicolumn{3}{c}{\textbf{Computational staining}} & \multicolumn{3}{c}{\textbf{Computational destaining}} \\
\textbf{Image} & \textbf{U\_C} & \textbf{U\_H} & \textbf{H$^{*}$\_C} & \textbf{H\_D} & \textbf{H\_U} & \textbf{D\_U$^{*}$} \\
\midrule
1 & -42 & -43 & 1 & 45 & 43 & -3 \\
2 & -30 & -26 & -4 & 37 & 26 & -11 \\
3 & -39 & -47 & 8 & 40 & 47 & 7 \\
4 & -48 & -49 & 1 & 48 & 49 & 1 \\
5 & -48 & -44 & -3 & 44 & 44 &0  \\
6 & -32 & -34 & 2 & 44 & 34 & -10 \\
7 & -19 & -21 & 1 & 32 & 21 & -12 \\
8 & -53 & -53 & 0 & 58 & 53 & -5 \\
9 & -45 & -48 & 3 & 56 & 48 & -8 \\
10 & -42 & -40 & -2 & 44 & 40 & -4 \\
11 & -43 & -42 & -2 & 48 & 42 & -6 \\
12 & -56 & -66 & 10 & 62 & 66 & 4 \\
13 & -50 & -58 & 8 & 56 & 58 & 2 \\
\textbf{MEAN} & \textbf{-42} & \textbf{-44} & \textbf{2} & \textbf{47} & \textbf{44} & \textbf{-3} \\
\bottomrule
\end{tabular}
\caption{Average pixel intensity differences following computational staining and destaining: Difference between Native nonstained and computationally stained (U\_C); native nonstained and H\&E dye stained (U\_H); ground truth H\&E dye stained and computationally stained (H*\_C); H\&E dye stained and computationally destained (H\_D); H\&E dye stained and native nonstained (H\_U); computationally destained and ground truth native nonstained (D\_U*). All values are in pixel intensities (0 to 255) calculated by subtracting the 2\textsuperscript{nd} from 1\textsuperscript{st} image. Positive values indicate decrease in average pixel intensities and negative values indicate gain. Values have been rounded to nearest integer. H is H\&E dye stained image, C is computationally stained image, D is computationally destained image and U is native nonstained image. Ground truth images are indicated with `*' to facilitate comparisons with computational images when necessary.}
\label{table:RGB_pixel_delta}
\end{table}

\newpage
\begin{table}
\centering
\renewcommand{\arraystretch}{0.6}
\begin{tabular}{cccccc}
\toprule
\textbf{Image} & \textbf{Any tumor} & \textbf{Healthy} & \textbf{G3} & \textbf{G4} & \textbf{G5} \\
\midrule
1   & 0.90      & 0.96      & 0.90      & -         & - \\
2   & 0.86      & 0.55      & -         & 0.78      & - \\
3   & -         & 1.00      & -         & -         & - \\
4   & 0.92      & 0.89      & 0.76      & -         & - \\
5   & 0.52      & 0.90      & -         & 0.49      & 0.64 \\
6   & 0.80      & 0.93      & 0.58      & -         & - \\
7   & 0.70      & 0.94      & 0.53      & -         & - \\
8   & 0.79      & 0.92      & -         & 0.77      & - \\
9   & 0.58      & 0.96      & 0.48      & -         & - \\
10  & 0.86      & 0.86      & 0.70      & 0.72      & - \\
11  & 0.92      & 0.99      & 0.92      & -         & - \\
12  & -         & 1.00      & -         & -         & - \\
13  & 0.93      & 0.78      & -         & 0.89      & - \\
\textbf{MEAN} & \textbf{0.79} & \textbf{0.90} & \textbf{0.70} & \textbf{0.73} & \textbf{0.64} \\
\bottomrule
\end{tabular}
\caption{Intersection over union (IoU) based agreement between pathologists for tumor signatures provided using computationally stained images compared with those using ground truth Hematoxylin and Eosin (H\&E) dye stained images. Any tumor: any tumor grade; Healthy: tissue without tumors; G3: Gleason grade 3; G4: Gleason grade 4; G5: Gleason grade 5. `\textbf{-}' indicates that the tumor (or tumor grade) was absent on a particular image. Higher IoU score is better, with a score of 1.0 representing perfect match of labels.}
\label{table:inter_rater_agreement}
\end{table}

\newpage

\begin{table}
\centering
\renewcommand{\arraystretch}{0.8}
\begin{tabular}{ccc}
\toprule
\textbf{Image} & \textbf{Initial diagnosis after biopsy} & \textbf{Diagnosis using computationally stained image} \\
\midrule
1 & 40\% grade 3 tumor in core & 40\% grade 3 tumor in core \\
2 & 50\% grade 3 tumor in core & 90\% grade 3 tumor in core $^{*, \dagger}$ \\
3 & Benign core & Benign core \\
4 & 50\% grade 3 tumor in core & 50\% grade 3 tumor (majority) with \\
& & traces of grade 4 tumor $^{\dagger}$ \\
5 & 50\% grade 4 and 5 tumor on core & 50\% grade 4 and 5 tumor on core \\
& (G4 $>$ G5) & (G4 $<$ G5) $^{\dagger}$ \\
6 & 40\% grade 3 and 4 tumor in core& 40\% grade 3 and 4 tumor in core\\
& (G3 $\gg$ G4) & (G3 $\gg$ G4) $^{\dagger}$ \\
7 & 40\% grade 3 tumor in core & 40\% grade 3 and 4 tumor in core \\
&  & (G4 $\gg$ G3) $^{\dagger}$ \\
8 & 40\% grade 3 and 4 tumor in core & 40\% grade 3 and 4 tumor in core \\
9 & 20\% grade 3 tumor in core & 20\% grade 3 tumor in core \\
10 & 90\% grade 3 and 4 tumor in core & 90\% grade 3 and 4 tumor in core \\
11 & Healthy core & Tiny focus of grade 3 tumor in core $^{*, \dagger}$\\
12 & Healthy core & Healthy core \\
13 & 90\% grade 4 tumor in core & 90\% grade 3 and 4 tumor in core \\
&  & (G3 $\ll$ G4) $^{\dagger}$ \\
\bottomrule
\end{tabular}%
\caption{Comparison of tumor grades between original expert microscopic diagnosis [as reported in the Electronic Health Records (EHR)] using the Hematoxylin and Eosin (H\&E) stained glass slide and the diagnosis of the computationally stained image. $^*$ Agreements confirmed upon re-review of the microscopic slide and additional supportive studies; $^\dagger$ Not clinically significant within the context of the patients known tumor.}
\end{table}

\clearpage
\section*{References}
\bibliographystyle{naturemag}
\bibliography{bibliography}

\pagenumbering{gobble}

\begin{figure}[!htb]
    \centering
    \includegraphics[width=0.9\linewidth]{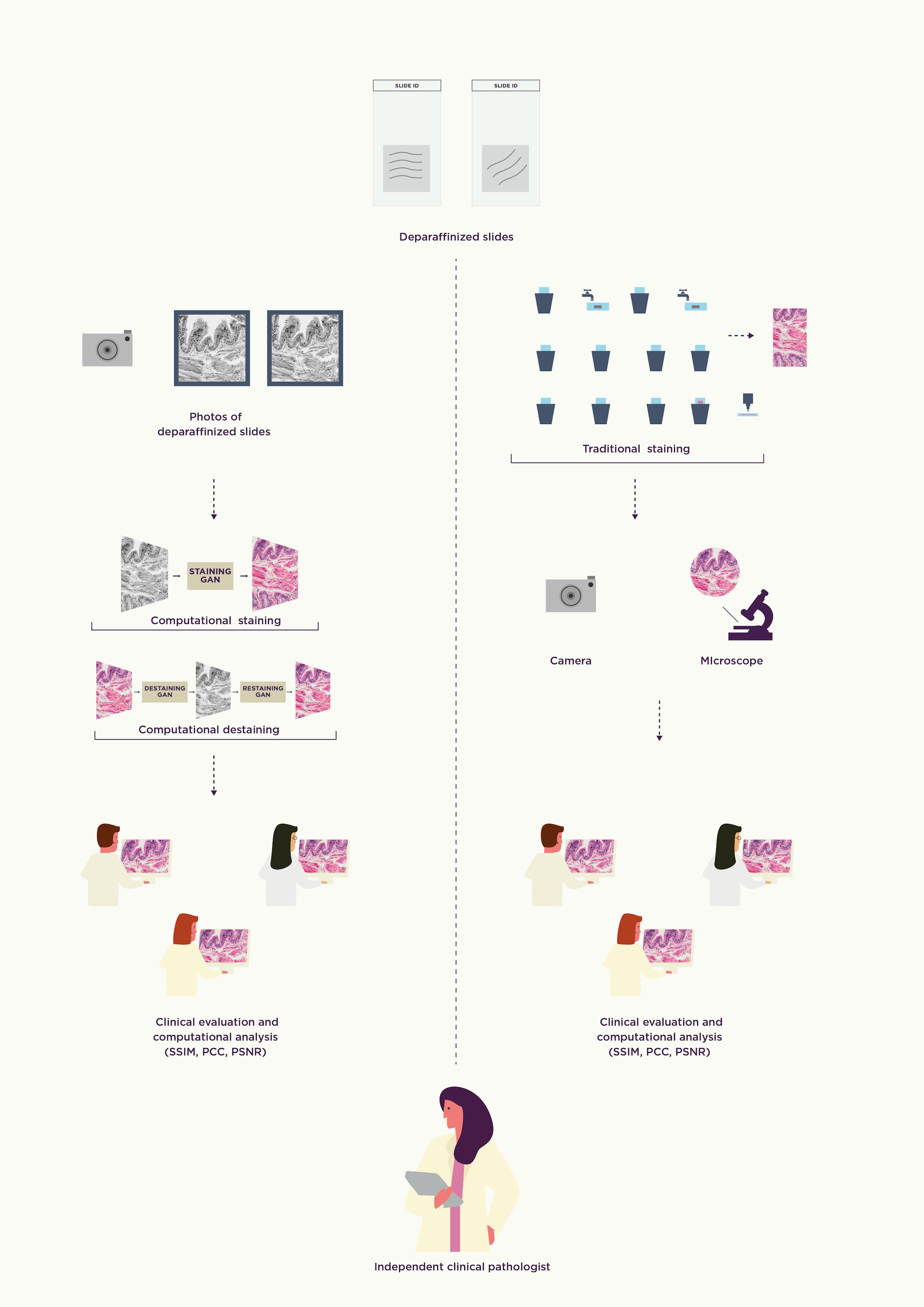}
    \caption{}
    \label{fig:process_comparison_figure}
\end{figure}
\newpage

\begin{figure}[!htbp]
\begin{center}

\begin{subfigure}[t]{0.15\linewidth}
	\begin{center}
	    I
	\end{center}
\end{subfigure}
\begin{subfigure}[t]{0.15\linewidth}
	\begin{center}
	    II
	\end{center}
\end{subfigure}
\begin{subfigure}[t]{0.15\linewidth}
	\begin{center}
	    III
	\end{center}
\end{subfigure}
\begin{subfigure}[t]{0.15\linewidth}
	\begin{center}
	    IV
	\end{center}
\end{subfigure}
\begin{subfigure}[t]{0.15\linewidth}
	\begin{center}
	    V
	\end{center}
\end{subfigure}

\begin{subfigure}[t]{0.15\linewidth}
	\makebox[0pt][r]{\makebox[30pt]{\raisebox{30pt}{(a)}}}\includegraphics[width=\linewidth]{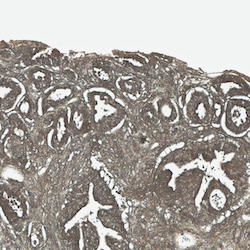}
\end{subfigure}
\begin{subfigure}[t]{0.15\linewidth}
	\includegraphics[width=\linewidth]{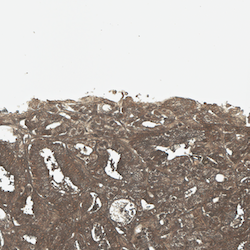}
\end{subfigure}
\begin{subfigure}[t]{0.15\linewidth}
	\includegraphics[width=\linewidth, angle=180, origin=c]{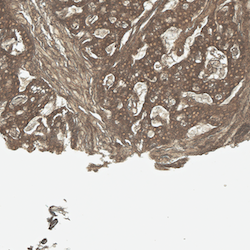}
\end{subfigure}
\begin{subfigure}[t]{0.15\linewidth}
	\includegraphics[width=\linewidth]{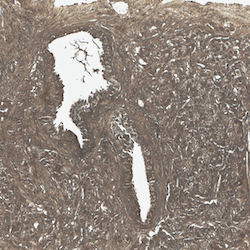}
\end{subfigure}
\begin{subfigure}[t]{0.15\linewidth}
	\includegraphics[width=\linewidth]{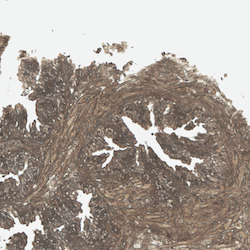}
\end{subfigure}

\begin{subfigure}[t]{0.15\linewidth}
	\makebox[0pt][r]{\makebox[30pt]{\raisebox{30pt}{(b)}}}\includegraphics[width=\linewidth]{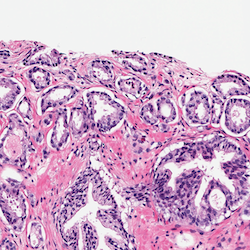}
\end{subfigure}
\begin{subfigure}[t]{0.15\linewidth}
	\includegraphics[width=\linewidth]{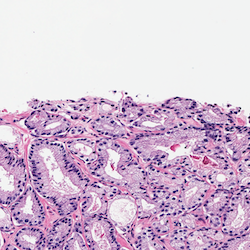}
\end{subfigure}
\begin{subfigure}[t]{0.15\linewidth}
	\includegraphics[width=\linewidth, angle=180, origin=c]{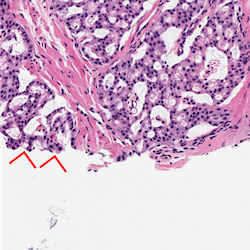}
\end{subfigure}
\begin{subfigure}[t]{0.15\linewidth}
	\includegraphics[width=\linewidth]{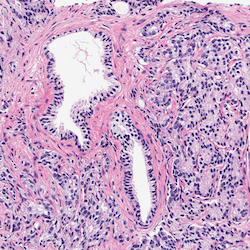}
\end{subfigure}
\begin{subfigure}[t]{0.15\linewidth}
	\includegraphics[width=\linewidth]{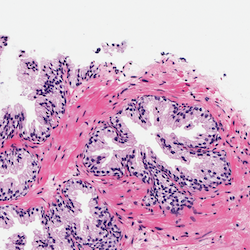}
\end{subfigure}

\begin{subfigure}[t]{0.15\linewidth}
	\makebox[0pt][r]{\makebox[30pt]{\raisebox{30pt}{(c)}}}\includegraphics[width=\linewidth]{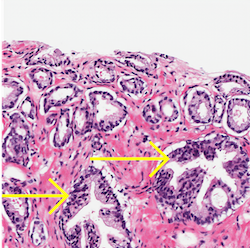}
\end{subfigure}
\begin{subfigure}[t]{0.15\linewidth}
	\includegraphics[width=\linewidth]{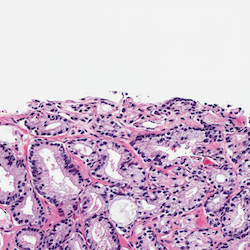}
\end{subfigure}
\begin{subfigure}[t]{0.15\linewidth}
	\includegraphics[width=\linewidth, angle=180, origin=c]{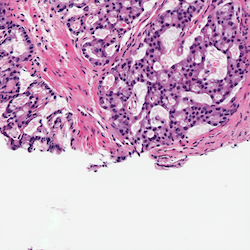}
\end{subfigure}
\begin{subfigure}[t]{0.15\linewidth}
	\includegraphics[width=\linewidth]{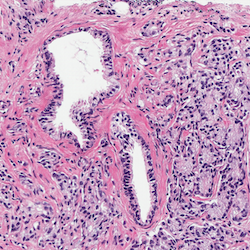}
\end{subfigure}
\begin{subfigure}[t]{0.15\linewidth}
	\includegraphics[width=\linewidth]{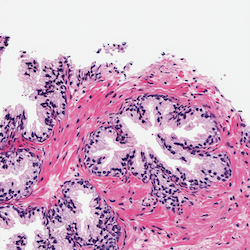}
\end{subfigure}

\begin{subfigure}[t]{0.15\linewidth}
	\makebox[0pt][r]{\makebox[30pt]{\raisebox{30pt}{(d)}}}\includegraphics[width=\linewidth]{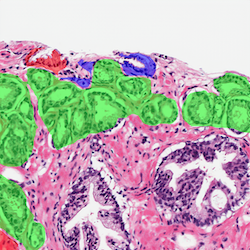}
\end{subfigure}
\begin{subfigure}[t]{0.15\linewidth}
	\includegraphics[width=\linewidth]{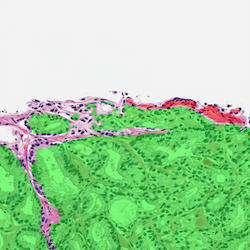}
\end{subfigure}
\begin{subfigure}[t]{0.15\linewidth}
	\includegraphics[width=\linewidth, angle=180, origin=c]{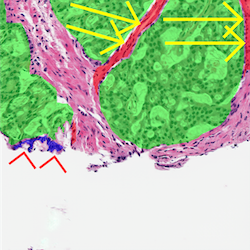}
\end{subfigure}
\begin{subfigure}[t]{0.15\linewidth}
	\includegraphics[width=\linewidth]{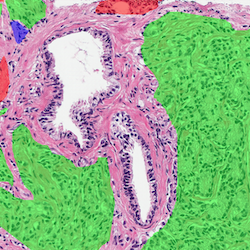}
\end{subfigure}
\begin{subfigure}[t]{0.15\linewidth}
	\includegraphics[width=\linewidth]{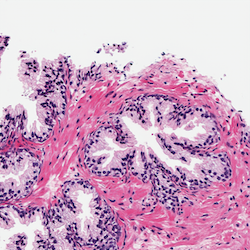}
\end{subfigure}

\end{center}

\begin{center}
\caption{ }
\label{fig:grid_figure}
\end{center}
\end{figure}
\newpage

\begin{figure}[!htbp]
\begin{center}
\begin{subfigure}[t]{0.9\linewidth}
	\begin{center}
	     \includegraphics[width=0.9\linewidth]{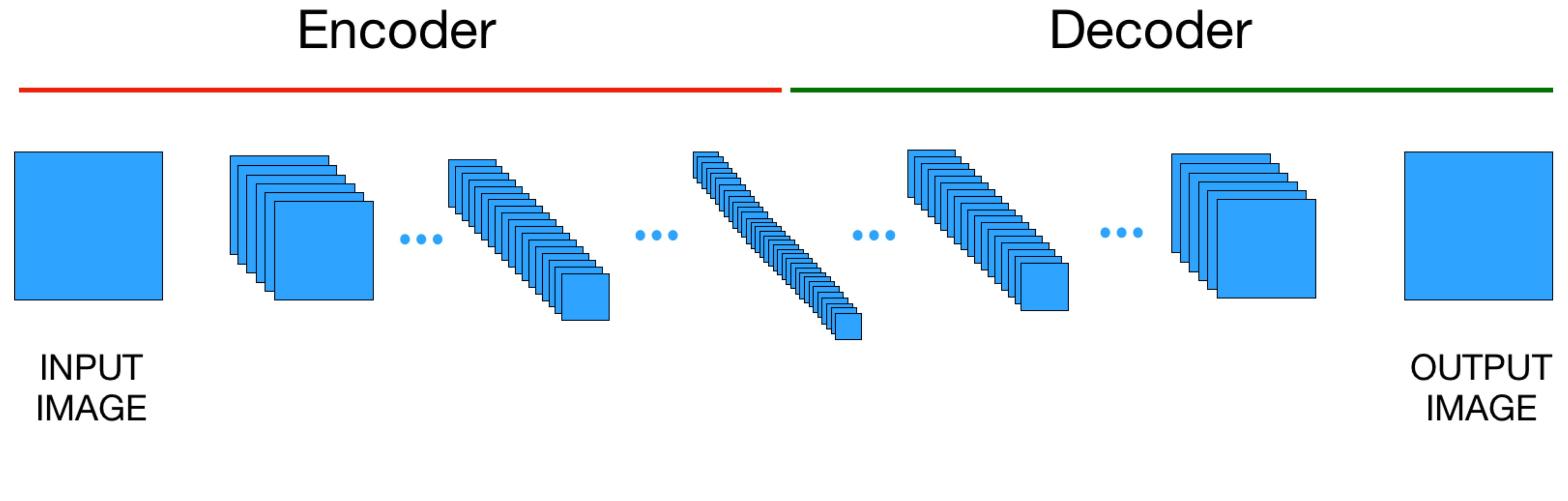}
	     \caption{}
	\end{center}
\end{subfigure}
\begin{subfigure}[t]{0.9\linewidth}
	\begin{center}
	     \includegraphics[width=0.9\linewidth]{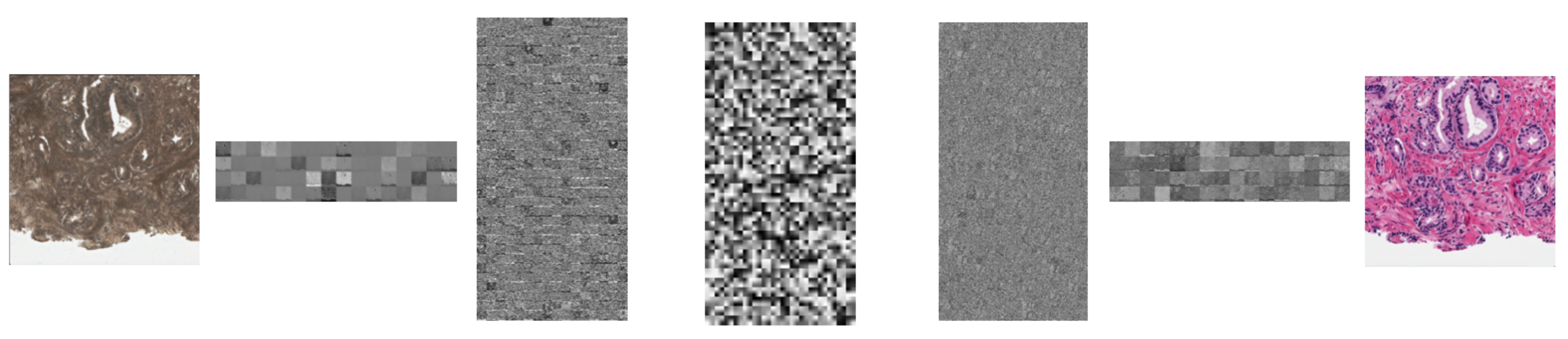}
	     \caption{}
	\end{center}
\end{subfigure}
\caption{ }
\label{fig:neural_network_layers_with_consolidated_activation_map}
\end{center}
\end{figure}
\newpage

\begin{figure}[!htbp]
\begin{center}

\begin{subfigure}[t]{0.13\linewidth}
	\begin{center}
	    I
	\end{center}
\end{subfigure}
\begin{subfigure}[t]{0.13\linewidth}
	\begin{center}
	    II
	\end{center}
\end{subfigure}
\begin{subfigure}[t]{0.13\linewidth}
	\begin{center}
	    III
	\end{center}
\end{subfigure}
\begin{subfigure}[t]{0.13\linewidth}
	\begin{center}
	    IV
	\end{center}
\end{subfigure}
\begin{subfigure}[t]{0.13\linewidth}
	\begin{center}
	    V
	\end{center}
\end{subfigure}

\begin{subfigure}[t]{0.13\linewidth}
	\makebox[0pt][r]{\makebox[30pt]{\raisebox{30pt}{(L1)}}}\includegraphics[width=\linewidth]{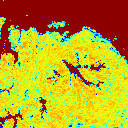}
\end{subfigure}
\begin{subfigure}[t]{0.13\linewidth}
	\includegraphics[width=\linewidth]{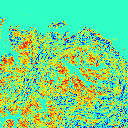}
\end{subfigure}
\begin{subfigure}[t]{0.13\linewidth}
	\includegraphics[width=\linewidth]{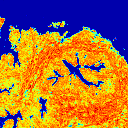}
\end{subfigure}
\begin{subfigure}[t]{0.13\linewidth}
	\includegraphics[width=\linewidth]{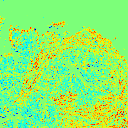}
\end{subfigure}
\begin{subfigure}[t]{0.13\linewidth}
	\includegraphics[width=\linewidth]{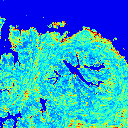}
\end{subfigure}

\begin{subfigure}[t]{0.13\linewidth}
	\makebox[0pt][r]{\makebox[30pt]{\raisebox{30pt}{(L2)}}}\includegraphics[width=\linewidth]{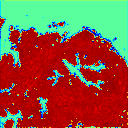}
\end{subfigure}
\begin{subfigure}[t]{0.13\linewidth}
	\includegraphics[width=\linewidth]{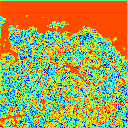}
\end{subfigure}
\begin{subfigure}[t]{0.13\linewidth}
	\includegraphics[width=\linewidth]{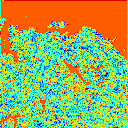}
\end{subfigure}
\begin{subfigure}[t]{0.13\linewidth}
	\includegraphics[width=\linewidth]{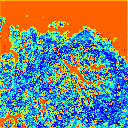}
\end{subfigure}
\begin{subfigure}[t]{0.13\linewidth}
	\includegraphics[width=\linewidth]{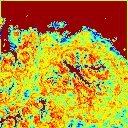}
\end{subfigure}

\begin{subfigure}[t]{0.13\linewidth}
	\makebox[0pt][r]{\makebox[30pt]{\raisebox{30pt}{(L3)}}}\includegraphics[width=\linewidth]{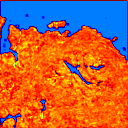}
\end{subfigure}
\begin{subfigure}[t]{0.13\linewidth}
	\includegraphics[width=\linewidth]{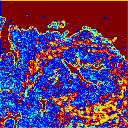}
\end{subfigure}
\begin{subfigure}[t]{0.13\linewidth}
	\includegraphics[width=\linewidth]{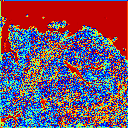}
\end{subfigure}
\begin{subfigure}[t]{0.13\linewidth}
	\includegraphics[width=\linewidth]{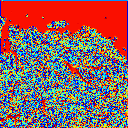}
\end{subfigure}
\begin{subfigure}[t]{0.13\linewidth}
	\includegraphics[width=\linewidth]{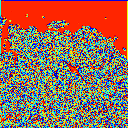}\end{subfigure}

\begin{subfigure}[t]{0.13\linewidth}
	\makebox[0pt][r]{\makebox[30pt]{\raisebox{30pt}{(L4)}}}\includegraphics[width=\linewidth]{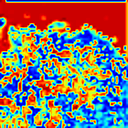}
\end{subfigure}
\begin{subfigure}[t]{0.13\linewidth}
	\includegraphics[width=\linewidth]{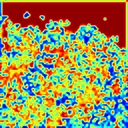}
\end{subfigure}
\begin{subfigure}[t]{0.13\linewidth}
	\includegraphics[width=\linewidth]{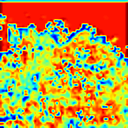}
\end{subfigure}
\begin{subfigure}[t]{0.13\linewidth}
	\includegraphics[width=\linewidth]{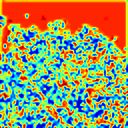}
\end{subfigure}
\begin{subfigure}[t]{0.13\linewidth}
	\includegraphics[width=\linewidth]{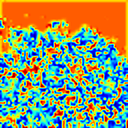}
\end{subfigure}

\begin{subfigure}[t]{0.13\linewidth}
	\makebox[0pt][r]{\makebox[30pt]{\raisebox{30pt}{(L5)}}}\includegraphics[width=\linewidth]{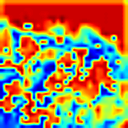}
\end{subfigure}
\begin{subfigure}[t]{0.13\linewidth}
	\includegraphics[width=\linewidth]{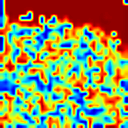}
\end{subfigure}
\begin{subfigure}[t]{0.13\linewidth}
	\includegraphics[width=\linewidth]{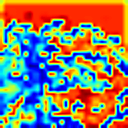}
\end{subfigure}
\begin{subfigure}[t]{0.13\linewidth}
	\includegraphics[width=\linewidth]{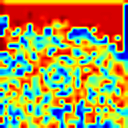}
\end{subfigure}
\begin{subfigure}[t]{0.13\linewidth}
	\includegraphics[width=\linewidth]{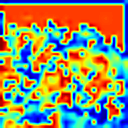}
\end{subfigure}

\begin{subfigure}[t]{0.13\linewidth}
	\makebox[0pt][r]{\makebox[30pt]{\raisebox{30pt}{(L16)}}}\includegraphics[width=\linewidth]{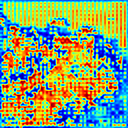}
\end{subfigure}
\begin{subfigure}[t]{0.13\linewidth}
	\includegraphics[width=\linewidth]{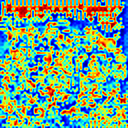}
\end{subfigure}
\begin{subfigure}[t]{0.13\linewidth}
	\includegraphics[width=\linewidth]{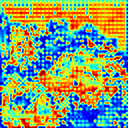}
\end{subfigure}
\begin{subfigure}[t]{0.13\linewidth}
	\includegraphics[width=\linewidth]{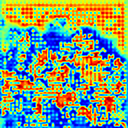}
\end{subfigure}
\begin{subfigure}[t]{0.13\linewidth}
	\includegraphics[width=\linewidth]{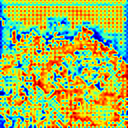}
\end{subfigure}

\begin{subfigure}[t]{0.13\linewidth}
	\makebox[0pt][r]{\makebox[30pt]{\raisebox{30pt}{(L17)}}}\includegraphics[width=\linewidth]{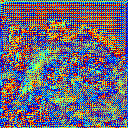}
\end{subfigure}
\begin{subfigure}[t]{0.13\linewidth}
	\includegraphics[width=\linewidth]{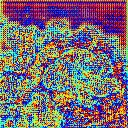}
\end{subfigure}
\begin{subfigure}[t]{0.13\linewidth}
	\includegraphics[width=\linewidth]{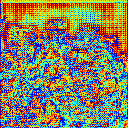}
\end{subfigure}
\begin{subfigure}[t]{0.13\linewidth}
	\includegraphics[width=\linewidth]{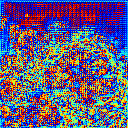}
\end{subfigure}
\begin{subfigure}[t]{0.13\linewidth}
	\includegraphics[width=\linewidth]{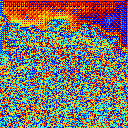}
\end{subfigure}

\begin{subfigure}[t]{0.13\linewidth}
	\makebox[0pt][r]{\makebox[30pt]{\raisebox{30pt}{(L18)}}}\includegraphics[width=\linewidth]{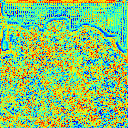}
\end{subfigure}
\begin{subfigure}[t]{0.13\linewidth}
	\includegraphics[width=\linewidth]{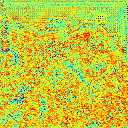}
\end{subfigure}
\begin{subfigure}[t]{0.13\linewidth}
	\includegraphics[width=\linewidth]{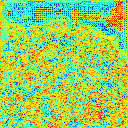}
\end{subfigure}
\begin{subfigure}[t]{0.13\linewidth}
	\includegraphics[width=\linewidth]{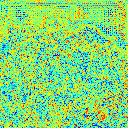}
\end{subfigure}
\begin{subfigure}[t]{0.13\linewidth}
	\includegraphics[width=\linewidth]{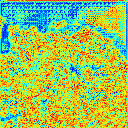}
\end{subfigure}

\begin{subfigure}[t]{0.13\linewidth}
	\makebox[0pt][r]{\makebox[30pt]{\raisebox{30pt}{(L19)}}}\includegraphics[width=\linewidth]{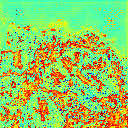}
\end{subfigure}
\begin{subfigure}[t]{0.13\linewidth}
	\includegraphics[width=\linewidth]{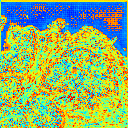}
\end{subfigure}
\begin{subfigure}[t]{0.13\linewidth}
	\includegraphics[width=\linewidth]{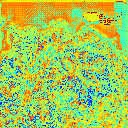}
\end{subfigure}
\begin{subfigure}[t]{0.13\linewidth}
	\includegraphics[width=\linewidth]{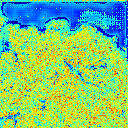}
\end{subfigure}
\begin{subfigure}[t]{0.13\linewidth}
	\includegraphics[width=\linewidth]{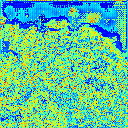}
\end{subfigure}

\begin{subfigure}[t]{\linewidth}
	\begin{center}
	\makebox[0pt][r]{\makebox[30pt]{\raisebox{30pt}{ }}}\includegraphics[width=0.60\linewidth]{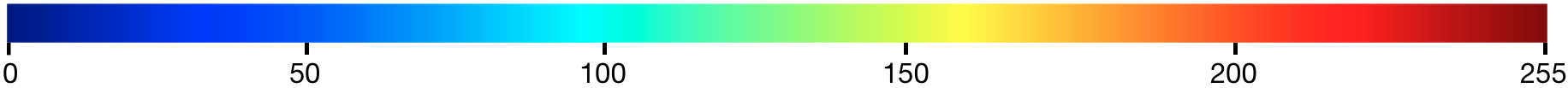}
	\end{center}
\end{subfigure}


\caption{ }
\label{fig:color_coded_activation_maps}
\end{center}
\end{figure}
\newpage

\begin{figure}[htbp]
\begin{center}

\begin{subfigure}[t]{0.13\linewidth}
	\begin{center}
	    I
	\end{center}
\end{subfigure}
\begin{subfigure}[t]{0.13\linewidth}
	\begin{center}
	    II
	\end{center}
\end{subfigure}
\begin{subfigure}[t]{0.13\linewidth}
	\begin{center}
	    III
	\end{center}
\end{subfigure}
\begin{subfigure}[t]{0.13\linewidth}
	\begin{center}
	    IV
	\end{center}
\end{subfigure}
\begin{subfigure}[t]{0.13\linewidth}
	\begin{center}
	    V
	\end{center}
\end{subfigure}

\begin{subfigure}[t]{0.13\linewidth}
	\makebox[0pt][r]{\makebox[30pt]{\raisebox{30pt}{(L1)}}}\includegraphics[width=\linewidth]{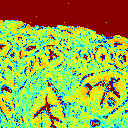}
\end{subfigure}
\begin{subfigure}[t]{0.13\linewidth}
	\includegraphics[width=\linewidth]{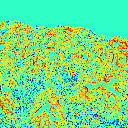}
\end{subfigure}
\begin{subfigure}[t]{0.13\linewidth}
	\includegraphics[width=\linewidth]{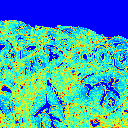}
\end{subfigure}
\begin{subfigure}[t]{0.13\linewidth}
	\includegraphics[width=\linewidth]{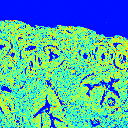}
\end{subfigure}
\begin{subfigure}[t]{0.13\linewidth}
	\includegraphics[width=\linewidth]{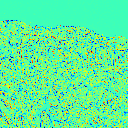}
\end{subfigure}

\begin{subfigure}[t]{0.13\linewidth}
	\makebox[0pt][r]{\makebox[30pt]{\raisebox{30pt}{(L2)}}}\includegraphics[width=\linewidth]{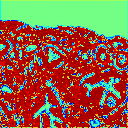}
\end{subfigure}
\begin{subfigure}[t]{0.13\linewidth}
	\includegraphics[width=\linewidth]{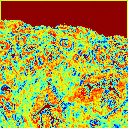}
\end{subfigure}
\begin{subfigure}[t]{0.13\linewidth}
	\includegraphics[width=\linewidth]{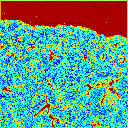}
\end{subfigure}
\begin{subfigure}[t]{0.13\linewidth}
	\includegraphics[width=\linewidth]{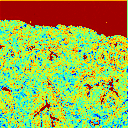}
\end{subfigure}
\begin{subfigure}[t]{0.13\linewidth}
	\includegraphics[width=\linewidth]{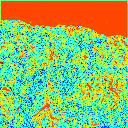}
\end{subfigure}

\begin{subfigure}[t]{0.13\linewidth}
	\makebox[0pt][r]{\makebox[30pt]{\raisebox{30pt}{(L3)}}}\includegraphics[width=\linewidth]{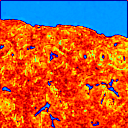}
\end{subfigure}
\begin{subfigure}[t]{0.13\linewidth}
	\includegraphics[width=\linewidth]{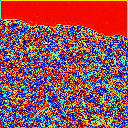}
\end{subfigure}
\begin{subfigure}[t]{0.13\linewidth}
	\includegraphics[width=\linewidth]{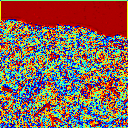}
\end{subfigure}
\begin{subfigure}[t]{0.13\linewidth}
	\includegraphics[width=\linewidth]{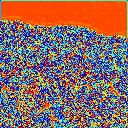}
\end{subfigure}
\begin{subfigure}[t]{0.13\linewidth}
	\includegraphics[width=\linewidth]{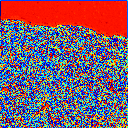}
\end{subfigure}

\begin{subfigure}[t]{0.13\linewidth}
	\makebox[0pt][r]{\makebox[30pt]{\raisebox{30pt}{(L4)}}}\includegraphics[width=\linewidth]{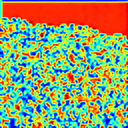}
\end{subfigure}
\begin{subfigure}[t]{0.13\linewidth}
	\includegraphics[width=\linewidth]{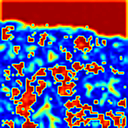}
\end{subfigure}
\begin{subfigure}[t]{0.13\linewidth}
	\includegraphics[width=\linewidth]{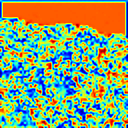}
\end{subfigure}
\begin{subfigure}[t]{0.13\linewidth}
	\includegraphics[width=\linewidth]{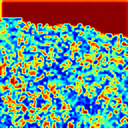}
\end{subfigure}
\begin{subfigure}[t]{0.13\linewidth}
	\includegraphics[width=\linewidth]{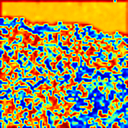}
\end{subfigure}

\begin{subfigure}[t]{0.13\linewidth}
	\makebox[0pt][r]{\makebox[30pt]{\raisebox{30pt}{(L5)}}}\includegraphics[width=\linewidth]{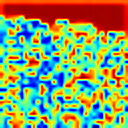}
\end{subfigure}
\begin{subfigure}[t]{0.13\linewidth}
	\includegraphics[width=\linewidth]{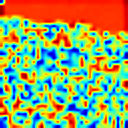}
\end{subfigure}
\begin{subfigure}[t]{0.13\linewidth}
	\includegraphics[width=\linewidth]{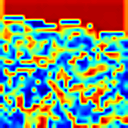}
\end{subfigure}
\begin{subfigure}[t]{0.13\linewidth}
	\includegraphics[width=\linewidth]{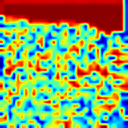}
\end{subfigure}
\begin{subfigure}[t]{0.13\linewidth}
	\includegraphics[width=\linewidth]{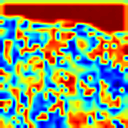}
\end{subfigure}

\begin{subfigure}[t]{0.13\linewidth}
	\makebox[0pt][r]{\makebox[30pt]{\raisebox{30pt}{(L16)}}}\includegraphics[width=\linewidth]{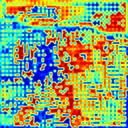}
\end{subfigure}
\begin{subfigure}[t]{0.13\linewidth}
	\includegraphics[width=\linewidth]{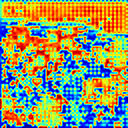}
\end{subfigure}
\begin{subfigure}[t]{0.13\linewidth}
	\includegraphics[width=\linewidth]{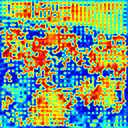}
\end{subfigure}
\begin{subfigure}[t]{0.13\linewidth}
	\includegraphics[width=\linewidth]{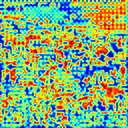}
\end{subfigure}
\begin{subfigure}[t]{0.13\linewidth}
	\includegraphics[width=\linewidth]{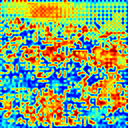}
\end{subfigure}

\begin{subfigure}[t]{0.13\linewidth}
	\makebox[0pt][r]{\makebox[30pt]{\raisebox{30pt}{(L17)}}}\includegraphics[width=\linewidth]{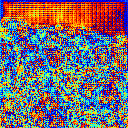}
\end{subfigure}
\begin{subfigure}[t]{0.13\linewidth}
	\includegraphics[width=\linewidth]{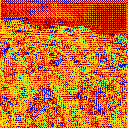}
\end{subfigure}
\begin{subfigure}[t]{0.13\linewidth}
	\includegraphics[width=\linewidth]{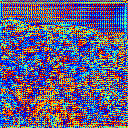}
\end{subfigure}
\begin{subfigure}[t]{0.13\linewidth}
	\includegraphics[width=\linewidth]{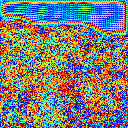}
\end{subfigure}
\begin{subfigure}[t]{0.13\linewidth}
	\includegraphics[width=\linewidth]{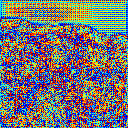}
\end{subfigure}

\begin{subfigure}[t]{0.13\linewidth}
	\makebox[0pt][r]{\makebox[30pt]{\raisebox{30pt}{(L18)}}}\includegraphics[width=\linewidth]{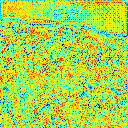}
\end{subfigure}
\begin{subfigure}[t]{0.13\linewidth}
	\includegraphics[width=\linewidth]{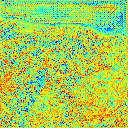}
\end{subfigure}
\begin{subfigure}[t]{0.13\linewidth}
	\includegraphics[width=\linewidth]{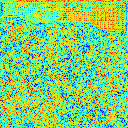}
\end{subfigure}
\begin{subfigure}[t]{0.13\linewidth}
	\includegraphics[width=\linewidth]{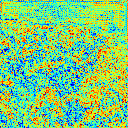}
\end{subfigure}
\begin{subfigure}[t]{0.13\linewidth}
	\includegraphics[width=\linewidth]{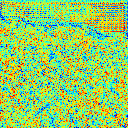}
\end{subfigure}

\begin{subfigure}[t]{0.13\linewidth}
	\makebox[0pt][r]{\makebox[30pt]{\raisebox{30pt}{(L19)}}}\includegraphics[width=\linewidth]{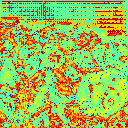}
\end{subfigure}
\begin{subfigure}[t]{0.13\linewidth}
	\includegraphics[width=\linewidth]{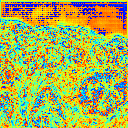}
\end{subfigure}
\begin{subfigure}[t]{0.13\linewidth}
	\includegraphics[width=\linewidth]{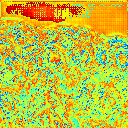}
\end{subfigure}
\begin{subfigure}[t]{0.13\linewidth}
	\includegraphics[width=\linewidth]{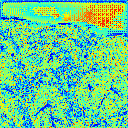}
\end{subfigure}
\begin{subfigure}[t]{0.13\linewidth}
	\includegraphics[width=\linewidth]{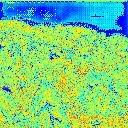}
\end{subfigure}

\begin{subfigure}[t]{\linewidth}
	\begin{center}
	\makebox[0pt][r]{\makebox[30pt]{\raisebox{30pt}{ }}}\includegraphics[width=0.60\linewidth]{figures/COLORBAR.png}
	\end{center}
\end{subfigure}

\end{center}

\begin{center}
\caption{ }
\label{fig:color_coded_activation_maps}
\end{center}
\end{figure}
\newpage

\begin{figure}[htbp]
\begin{center}
\begin{subfigure}[t]{0.9\linewidth}
	\begin{center}
	    \includegraphics[width=0.9\linewidth]{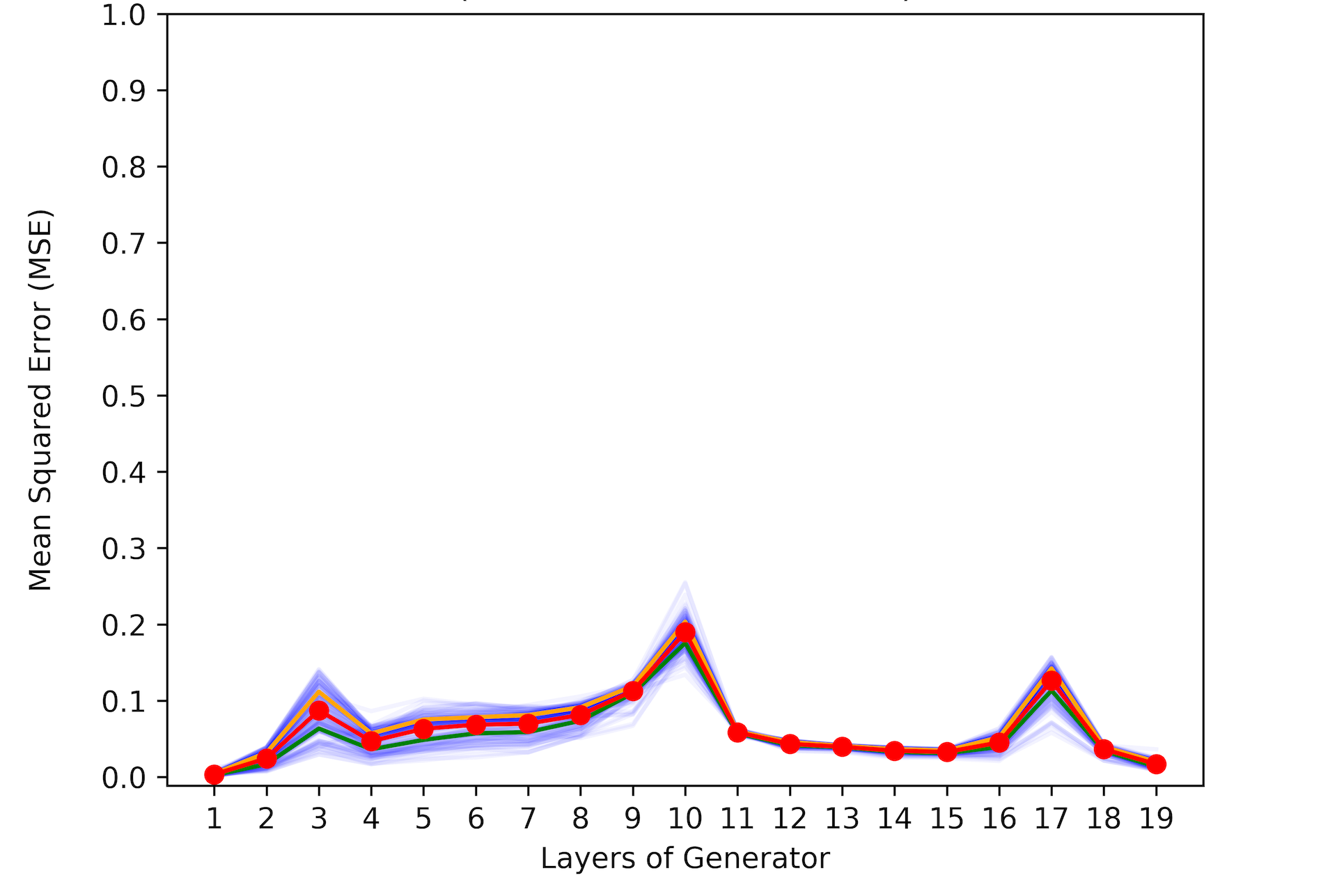}
	    \caption{}
	\end{center}
\end{subfigure}
\begin{subfigure}[t]{0.9\linewidth}
	\begin{center}
	    \includegraphics[width=0.9\linewidth]{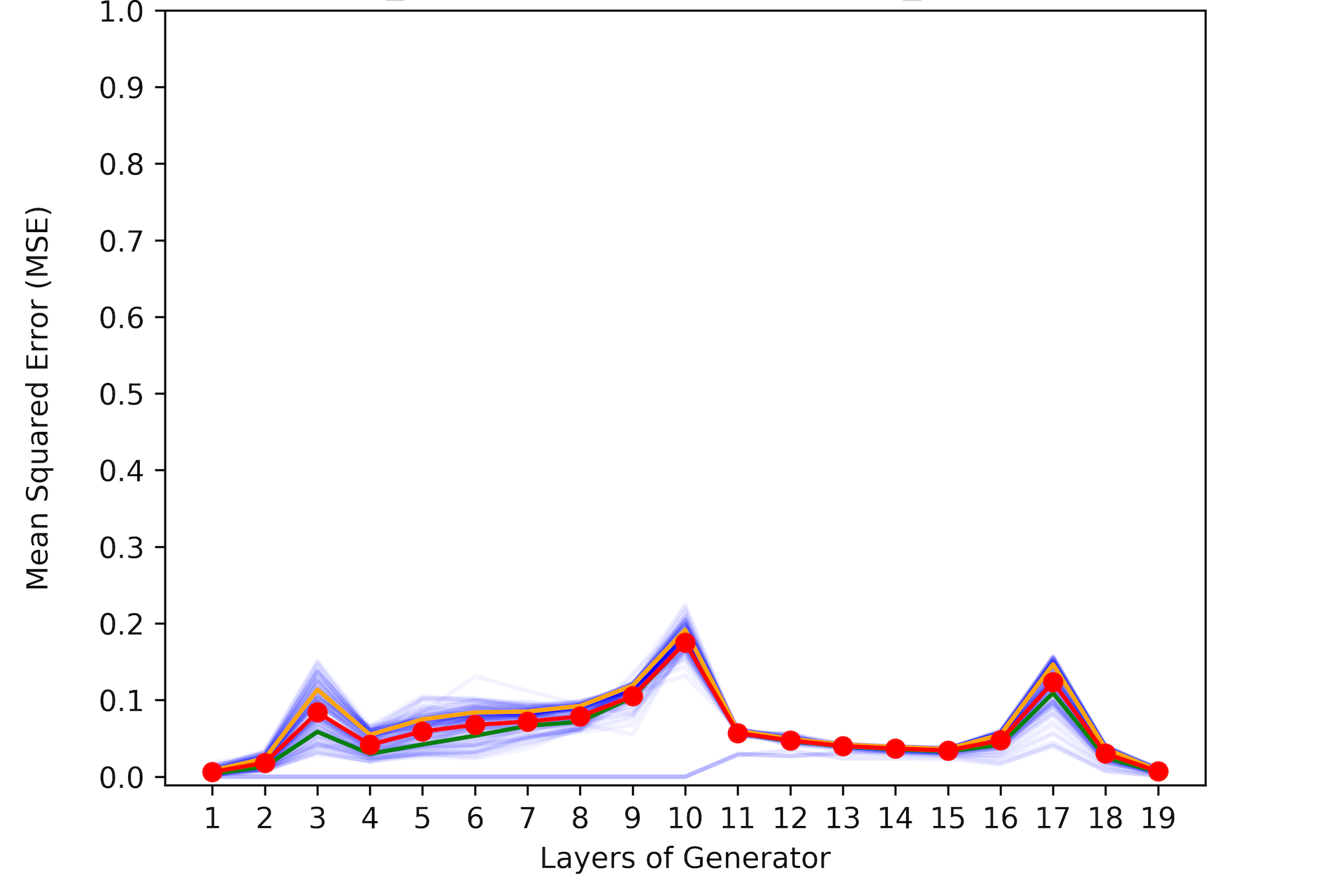}
	    \caption{}
	\end{center}
\end{subfigure}
\caption{ }
\label{figure:activation_mse}
\end{center}
\end{figure}

\end{document}